\documentstyle[12pt,a4wide]{article}
\title{\mbox{\LARGE{Primordial Magnetic Fields from}}\\
\mbox{\LARGE{Superconducting Cosmic Strings}}} 
\author{
\mbox{\large{Konstantinos Dimopoulos}}
\thanks{e-mail: K.Dimopoulos@damtp.cam.ac.uk}
\vspace{0.4cm}\\
\mbox{\normalsize{\em Department of Applied Mathematics and
Theoretical Physics,}}\\
\mbox{\normalsize{\em University of Cambridge, Silver Street,}}\\
\mbox{\normalsize{\em Cambridge, CB3 9EW, U.K.}}
}

\begin{document}
\begin{titlepage}
\maketitle
\begin{abstract}

Cosmic strings are stable topological defects that may have been
created at a phase transition in the early universe. It is a growing
belief that, for a wide range of theoretical models, such strings may
be superconducting and carry substantial currents which have
important astrophysical and cosmological effects. This paper explores 
the possibility of generation of a primordial magnetic field by
a network of charged--current carrying cosmic strings. The field is
created by vorticity, generated in the primordial plasma due to the
strings' motion and gravitational pull. In the case of superconducting
strings formed at the breaking of grand unification, it is found that
strong magnetic fields of high coherence can be generated in that
way. Such fields could account for the observed galactic and
intergalactic magnetic fields since they suffice to seed magnetic
dynamos on galactic scales.  

\end{abstract}
\vspace{1cm}
\flushright{DAMTP-97-30}
\end{titlepage}

\section{Introduction}

The origin of the observed galactic magnetic fields remains elusive. 
The magnetic field of the Milky Way and the nearby galaxies is of the
order of a $\mu Gauss$ and is coherent over $kpc$ scales. In the Milky
Way the magnetic field is orientated along the spiral arms alternating
its direction from one arm to the other. This strongly suggests that
the field is supported by a dynamo mechanism which arranges it along
the spiral density waves \cite{parker}\cite{dwaves}. Similar evidence for
a dynamo mechanism comes from galaxies at a stage of rapid star
formation (so called star-burst galaxies \cite{kron}) where the
magnetic flux needs to be amplified by a factor of \mbox{$\sim 5$} to
account for the observations \cite{parker}.

It is, therefore, a wide belief that the galactic magnetic fields are
generated through a dynamo mechanism \cite{dynamo}\cite{zeld}. A
number of mean field theory dynamo models exist in the literature, the
most popular of which is the well known $\alpha-\Omega$ dynamo. The
basic idea of the dynamo mechanism is that a weak seed field could be
amplified by the turbulent motion of ionised gas, which follows the
differential rotation of the galaxy. The growth of the field is
exponential and, thus, its strength can be increased several orders of
magnitude in  only a few e-foldings of amplification. When the field
reaches the equipartition energy (\mbox{$\sim\mu Gauss$}) then its
growth is suppressed by dynamical back-reaction. 

If the time scale of growth of the field is no more than a galactic
rotation period \mbox{$\sim 10^{8}yrs$}, then the field amplification
factor since the collapse of the protogalaxy is of the order
\mbox{$\sim 10^{13}$}, given that the total number of galactic
rotations is \mbox{$\simeq 30$}. Thus, the seed field required has to
be at least \mbox{$\sim 10^{-20}Gauss$} on the comoving scale of a
protogalaxy (\mbox{$\sim 100\,kpc$}) at the time 
\mbox{$t_{gc}\sim 10^{15}sec$} of gravitational collapse. 
Since the collapse of the galaxies enhances their frozen-in magnetic
field by a factor of \mbox{$(\rho_{g}/\rho_{c})^{2/3}\sim 10^{3}$}
(where \mbox{$\rho_{g}\sim 10^{-24}g\,cm^{-3}$} is the typical mass
density of a galaxy and \mbox{$\rho_{c}\simeq 2\times
10^{-29}\Omega h^{2}\,g\,cm^{-3}$} is the current cosmic mass density),
the above seed field corresponds to a field of the order of
\mbox{$\sim10^{-23}Gauss$} over the comoving scale of \mbox{$\sim
1\,M\!pc$}. Assuming that the rms field scales as $a^{-2}$
with the expansion of the universe ($a$ is the
scale factor of the universe, \mbox{$a\propto t^{2/3}$} in the matter
era) we find that the magnitude of the seed field at the start of
structure formation has to be at least \mbox{$\sim
10^{-22}Gauss\times(t_{gc}/t_{eq})^{4/3}$} \mbox{$\sim
10^{-21}Gauss$}, where \mbox{$t_{eq}\sim 10^{11}sec$} is the time of
equal matter and radiation energy densities. 

The origin and nature of this seed field is still an open
question. Many authors have argued that it could be produced by
stellar winds and other explosions \cite{zeld} but such a field would
be extremely incoherent over galactic scales. Significant
incoherences of the field have been shown \cite{kuls} to destabilise
and destroy the galactic dynamo and, therefore, it seems that
coherency is a crucial factor for seeding the galactic magnetic
fields. Also, the existing evidence for intergalactic fields
\cite{kron}\cite{rees} led people to believe that the
origin of galactic seed fields may be truly primordial. 

Most of the attempts to create a primordial magnetic field in the
early universe involve either inflation or phase transitions
because the creation of a primordial field can occur only in out of
thermal equilibrium conditions \cite{sacha}. 

At phase transitions
primordial magnetic fields can be created on the surface of the bubble
walls if the transition is a first order one \cite{bubbles}, or due to
stochastic Higgs--field gradients \cite{higgs}. Unfortunately though,
phase transitions occur very early and the causal horizon is of much
smaller comoving scale than the protogalactic one (e.g. for the
electroweak transition the comoving scale of the horizon is
\mbox{$\sim 10^{-3}pc$}). Thus, the generated magnetic field is too
incoherent to successfully trigger the galactic dynamo. 

Due to this fact, inflation has been considered as another option that
could increase the coherency of the created magnetic field \cite{infl}. 
However, the conformal invariance of electromagnetism suggested that
the field would scale as $a^{-2}$ during the inflationary period
which, as a result, diminished the field strength to much lower values
than the required seed field limit. In order to overcome this problem,
additional terms have been introduced in the Lagrangian to break
explicitly conformal invariance. Even so, most attempts produced too
weak seed fields.

Other attempts to generate an adequate seed field involve string
theory cosmology \cite{stth}. The problem has been pushed even further
back in time, into a possible pre-big bang epoch of negative
time, where dilaton inflation can take place. These models, although
attractive, suffer from lack of understanding of the interface between
the dilaton inflation era and the usual, post-big bang radiation
era. 

Perhaps the most realistic approach to the problem has been the cosmic
string scenario, where the magnetic field is generated by vortical
motions inside the wakes of cosmic strings \cite{cosst}. Vortical
generation of a magnetic field has been an early idea of Harrison
\cite{hary}, who considered the field to be created during the
radiation era inside expanding spinning volumes of plasma
(eddies). However, Rees has shown that Harrison's eddies would be
unstable and decay with cosmic expansion, whereas irrotational density
perturbations (lumps) from curvature fluctuations would grow and
dominate \cite{rees}. Rees suggested a different version of vortical
magnetic field growth which is similar to Harrison's idea but can be
applied to a gravitationally bound spinning body.

In the cosmic string scenario vorticity is generated in the wakes of
cosmic strings after structure formation begins. Therefore, the
vortical eddies are gravitationally bound and do not suffer from
instabilities. Matter in the trail of a string is substantially
ionised and so the Harrison--Rees mechanism can still operate.
The scale of coherency of the generated magnetic field is set by the
scale of wiggles on the string and, for wakes created at $t_{eq}$ it
can be up to $100\,kpc$. The field strength is of the order of
\mbox{$\sim 10^{-18}Gauss$}. Thus, cosmic strings are able to generate
magnetic fields of enough strength and coherency to seed the galactic
dynamo mechanism.

It is not very clear, though, whether stable vortical motions can be
generated by the rapid, stochastic, motion of the string wiggles. An
alternative mechanism by Avelino and Shellard \cite{paper} has
overcome this problem by considering dynamical friction. 
In this model, vorticity is generated not by the
wiggles but by the strings themselves, which drag matter behind them
and introduce circular motions over inter-string scales. The
magnetic field obtained though, is weak \mbox{$\sim 10^{-23}Gauss$}
and can only marginally seed the galactic dynamo. Also its coherency
is very high $\sim 100\,M\!pc$, which may be incompatible with
intergalactic field observations. 

In this paper we employ the mechanism of Avelino and Shellard in the
context of superconducting strings. It is an increasing belief that
cosmic strings may be generically superconducting and carry
substantial currents \cite{anne}. Charged current--carrying string
networks may evolve in a much different way that the usual,
non-superconducting case. As shown by Dimopoulos and Davis
\cite{mine}, superconducting networks may be more tangled with slower
moving strings. In this case we show that the magnetic field
generated can be much stronger than in the non-superconducting case
and still be coherent over protogalactic scales $\sim\,1\,M\!pc$. 

The structure of this paper is as follows. In Section 2 we give a
detailed overview of the field theory of string superconductivity
in both the bosonic and fermionic case. In Section 3 we deal with the
energy--momentum tensor of the superconducting string, which we use to
describe the string spacetime and its consequences on particle
deflection in Section 4. In Section 5 we calculate the primordial
magnetic field generated. First we find the momentum transfer from
the string to the plasma and the resulting rotational velocity of the
plasma vortical motions. Then, we estimate the generated primordial
magnetic field at the time when structure formation begins, taking
also into account constraints coming from the observations of the
microwave background anisotropy. Finally, in Section 6 we dicuss our
results and give our conclusions. Throughout this paper, unless stated
otherwise, we use natural units (\mbox{$\hbar=c=1$}) for which the
Planck mass is \mbox{$m_{P}=1.22\times 10^{19}GeV$}.

\section{String superconductivity}

String superconductivity was initially conceived by Witten
\cite{witt}. There are two types of models that give rise to
superconducting strings depending on whether the current carriers are
bosons or fermions. However, through a specific formalism, there is
exact correspondence between the two cases in most aspects. In what
follows, we give a description of bosonic and fermionic
superconductivity and develop the general formalism to treat them both.

\subsection{Bosonic superconductivity}

We will describe bosonic superconducting strings in the context of a
\mbox{$\overline{U(1)}\times U(1)_{em}$} theory as in 
\cite{witt}\ldots\cite{peter}.
This theory involves two scalar fields: The $\overline{U(1)}$ Higgs--field
$\phi$, which is responsible for the formation of the vortex 
and the $U(1)_{em}$ field $\sigma$ which breaks
electromagnetism inside the string and turns it superconducting. The
Lagrangian density of the theory is, 

\begin{equation}
{\cal L}= 
-\frac{1}{4}F_{\mu\nu}F^{\mu\nu}-\frac{1}{4}G_{\mu\nu}G^{\mu\nu}
+ (D_{\mu}\sigma)^{*}D^{\mu}\sigma + (D_{\mu}\phi)^{*}D^{\mu}\phi 
- V(\phi,\sigma)
\label{Lbos}
\end{equation}
where \mbox{$F_{\mu\nu}=\partial_{\mu}A_{\nu}-\partial_{\nu}A_{\mu}$}
and \mbox{$G_{\mu\nu}=\partial_{\mu}W_{\nu}-\partial_{\nu}W_{\mu}$},
 and also \mbox{$D_{\mu}\sigma=(\partial_{\mu}+ieA_{\mu})\sigma$} and 
\mbox{$D_{\mu}\phi=(\partial_{\mu}+igW_{\mu})\phi$} with $W_{\mu}$ and
$A_{\mu}$ being the gauge fields coupled to the vortex and the photon
field respectively ($e$ and $g$ are the relevant gauge couplings). 
The potential $V(\phi,\sigma)$ is given by,

\begin{equation}
V(\phi,\sigma)=\frac{1}{4}\lambda(\phi^{2}-\eta^{2})^{2}+
\frac{1}{4}\lambda'\sigma^{4}+f\sigma^{2}\phi^{2}-m^{2}\sigma^{2}
\label{V}
\end{equation}
where \mbox{$\lambda,\lambda',f\leq 1$} are coupling constants and
$\eta$ is the energy scale of the string. 

Through the above potential it is possible to form a string by
breaking $\overline{U(1)}$. For suitable values of parameters the coupling
between the scalar fields may force the breaking of electromagnetism
inside the string, making it superconducting.

One can rewrite the above potential as,

\begin{equation}
V(\phi,\sigma)=\frac{1}{4}\lambda(\phi^{2}-\eta^{2})^{2}+
\frac{1}{4}\lambda'(\sigma^{2}-\sigma_{0}^{2})^{2}+
f\sigma^{2}\phi^{2}-m^{2}\sigma^{2}
\end{equation}
where \mbox{$\sigma_{0}=\sqrt{2/\lambda'}\,m$} is the expectation
value of $\sigma$ inside the string core.

From the above it follows that, in order for $\overline{U(1)}$ to be
broken and to form the string we require, 
\mbox{$V(\eta,0)<V(0,\sigma_{0})
\Rightarrow\lambda\eta^{4}>\lambda'\sigma_{0}^{4}$} which 
gives the constraint,

\begin{equation}
(\frac{m}{\eta})^{4}<\frac{\lambda\lambda'}{4}
\label{A}
\end{equation}

Furthermore, in order for electromagnetism not to be also broken
outside the string we require that the unbroken state is a global
minimum of $V$, i.e. that
\mbox{$\frac{\partial^{2}}{\partial\sigma^{2}}V(\eta,0)>0$} 
which suggests that,

\begin{equation}
\frac{m^{2}}{f\eta^{2}}<1
\label{B}
\end{equation}

In order for electromagnetism to be broken inside the string we
require that the minimum of the potential in the core corresponds to a
non-zero value of $\sigma$, i.e. 
\mbox{$\frac{\partial^{2}}{\partial\sigma^{2}}V(0,\sigma\neq 0)>0$}
which gives, 

\begin{equation}
m^{2}>0
\end{equation}

Finally, in order for the $\sigma$-condensate to be contained inside
the string we require,

\begin{equation}
m_{\phi}^{-1}>(\sqrt{\lambda'}\,\sigma_{0})^{-1}\Rightarrow
(\frac{m}{\eta})^{2}>\frac{\lambda}{2}
\label{C}
\end{equation}
where \mbox{$m_{\phi}=\sqrt{\lambda}\,\eta$} is the mass of the Higgs
particle. The mass of the $\sigma$ particle is easily found as, 

\begin{equation}
m_{\sigma}^{2}=f\eta^{2}-m^{2}>0
\end{equation}
where we have also used (\ref{B}). 

By combining (\ref{A}) and (\ref{B}) we obtain,

\begin{equation}
\lambda<\frac{4}{\lambda}(\frac{m}{\eta})^{4}<\lambda'
\end{equation}

Under the above conditions the theory admits a vortex solution with
electromagnetism broken in its core. The $\sigma$ field may be
parametrised as,

\begin{equation}
\sigma=\sigma(r)\,e^{i\psi(z,t)}
\label{sigma}
\end{equation}
where we have assumed that the vortex core lies on the $z$-axis. The
phase field $\psi$ may vary randomly along the string and wind up in a
non-trivial way. This winding gives rise to the string current.

If one defines the current as,
\mbox{$
J_{\mu}=\frac{\partial{\cal L}}{\partial A_{\mu}}=
-ie\,[\sigma^{*}D_{\mu}\sigma-\sigma(D_{\mu}\sigma)^{*}]
$}
it is straightforward to show that the string current is,
\mbox{$
J_{\mu}=2e\sigma(r)^{2}(\partial_{\mu}\psi+eA_{\mu})
$}. Integrating over the string core gives,

\begin{equation}
J_{a}=2Ke\,(\partial_{a}\psi+eA_{a})
\label{J}
\end{equation}
where the constant $K$ is given by,

\begin{equation}
K\equiv\int d^{2}r\,\sigma^{2}\simeq\frac{1}{\lambda'}
\label{K}
\end{equation}

Varying (\ref{Lbos}) with respect to
$\phi$,$\sigma$,$W_{\mu}$,$A_{\mu}$ and $\psi$ respectively, gives the
following equations,

\begin{eqnarray}
 & & \Box\phi-gW^{\mu}W_{\mu}\phi-
\frac{1}{2}\lambda(\phi^{2}-\eta^{2})\phi-f\sigma^{2}\phi=0\\
 & & \nonumber\\
 & & \Box\sigma-(f\phi^{2}-m^{2})\sigma-
(\partial_{\mu}\psi+eA_{\mu})(\partial^{\mu}\psi+eA^{\mu})\sigma-
\frac{1}{2}\lambda'\sigma^{3}=0\\
 & & \nonumber\\
 & & \partial_{\mu}G^{\mu\nu}=g^{2}\phi^{2}W^{\nu}\\
 & & \nonumber\\
 & & \partial_{\mu}F^{\mu\nu}=J^{\nu}\\
 & & \nonumber\\
 & & \partial_{\mu}J^{\mu}=0\label{Jcons}
\end{eqnarray}
where \mbox{$\Box\equiv\partial_{\mu}\partial^{\mu}$}.

The last equation of the above expresses the dynamical conservation of
the string current. 

\subsection{Fermionic superconductivity}

Following Witten \cite{witt} we introduce two fermion fields $\Psi_{L}$ and
$\Psi_{R}$ which couple to the vortex field $\phi$ through a coupling
$\lambda$. Then, the Lagrangian density may be written as \cite{book},

\begin{equation}
{\cal L}=\overline{\Psi}_{L}i\!\not\!\!D\Psi_{L}+
\overline{\Psi}_{R}i\!\not\!\!D\Psi_{R}-
\lambda(\phi\overline{\Psi}_{L}\Psi_{R}+h.c.)
\label{Lferm}
\end{equation}
where \mbox{$\not\!\!D\equiv\gamma^{\mu}D_{\mu}$} and 
\mbox{$D_{\mu}\Psi =(\partial_{\mu}+iqA_{\mu})\Psi$} with
$\gamma^{\mu}$ being the Dirac matrices and $q$ the gauge coupling. 

As shown in \cite{witt} fermionic superconductivity resembles strongly
the bosonic one apart from some minor aspects such as particle
production. There is indeed a way to switch from one picture to the
other. Let us introduce the scalar field $y$ such that,

\begin{equation}
\overline{\Psi}\gamma^{a}\Psi\equiv
\frac{1}{\sqrt{\pi}}\,\varepsilon^{ab}\partial_{a}y
\label{Y}
\end{equation}
where $\varepsilon^{ab}$ is the 2-dimensional Levi-Civita tensor. 

Then, with the equivalence, \mbox{$q\equiv -\sqrt{2\pi K}e$} the
string current is,

\begin{equation}
J^{a}\equiv -q\overline{\Psi}\gamma^{a}\Psi=
\sqrt{2K}e\,\varepsilon^{ab}\partial_{a}y
\label{Jy}
\end{equation}

If we define $y$ such that,

\begin{equation}
\partial^{a}y=\sqrt{2K}\varepsilon^{ab}(\partial_{b}\psi+eA_{b})
\label{y}
\end{equation}
then the current in (\ref{Jy}) is identified with the one given in
(\ref{J}). As shown in \cite{witt} the above
definition of $y$ is consistent. In terms of this formalism we can
describe both the fermionic and the bosonic superconducting strings.

\subsection{The effective action}

Non-superconducting cosmic strings are described by the well--known
Goto--Nambu action \cite{book},

\begin{equation}
S=-\mu\int d^{2}\xi\sqrt{-\gamma}
\label{Sns}
\end{equation}
where $\mu$ is the energy per unit length of the string, $\xi^{a}$ are
the string world-sheet coordinates and $\gamma$ is the determinant of
the world-sheet metric
\mbox{$\gamma_{ab}=\partial_{a}x^{\mu}\partial_{b}x_{\mu}$}. The
relation between the spacetime and the world-sheet coordinates may be
expressed as,

\begin{equation}
x^{\mu}=x^{\mu}(\xi^{a})+n_{A}^{\mu}r^{A}
\end{equation}
where $n_{A}^{\mu}$ are vectors perpendicular to the string, that form
a basis for the 2-dimensional perpendicular space spanned by the
$r^{A}$ coordinates for which, \mbox{$r^{2}=r^{A}r_{A}$}.\footnote{We
will use $a,b,c$ e.t.c. indeces to 
denote the components on the 2-dimensional world-sheet of the string,
whereas $\lambda,\mu,\nu,$ e.t.c. indeces will be used to denote the
components in 4-dimensions. The projection of a vector $V^{\mu}$ on the
string world-sheet is \mbox{$V^{a}=(\partial^{a}\!x_{\mu})V^{\mu}$} where
$x^{\mu}$ are the 4-dimensional space coordinates. In a similar way we
can project higher rank tensors. Finally, the $A,B,C$ e.t.c. indeces
will be used for the 2-dimensional space perpendicular to the string.} 
The relation between the string metric $\gamma_{ab}$ and the
spacetime metric $g_{\mu\nu}$ to first order is \cite{turok},

\begin{equation}
g_{\mu\nu}\simeq\left( 
\begin{array}{lr}
\gamma_{ab} & 0 \\
 & \\
0 & \delta_{AB} 
\end{array}
\right)
\end{equation}

In the case of superconducting string the action has to account for
the existence of a current. In the bosonic picture this, in fact, is
equivalent with considering the addition of the $\sigma$ field
kinetic--energy 
term,\footnote{The potential energy due to the $\sigma$ field is
included into $\mu$.}

\begin{eqnarray}
\Delta S_{J} & = & \int d^{2}\xi\,d^{2}r\sqrt{-\gamma}\,|D_{a}\sigma|^{2}=
\int d^{2}r\,\sigma(r)^{2}\int 
d^{2}\xi\sqrt{-\gamma}\,|\partial_{a}\psi+eA_{a}|^{2}=\nonumber\\
 & & \nonumber\\
 & = & -\frac{1}{2}\int d^{2}\xi\sqrt{-\gamma}\,(\partial y)^{2}
\label{DSJ}
\end{eqnarray}
where \mbox{$(\partial y)^{2}\equiv (\partial_{a}y)(\partial^{a}y)$}
and we have used (\ref{sigma}), (\ref{K}) and (\ref{y}). 

Thus, with the addition of the current term the Goto--Nambu action
becomes \cite{turok}, 

\begin{equation}
\Delta S=-\int d^{2}\xi\sqrt{-\gamma}\,[\mu+\frac{1}{2}(\partial
y)^{2}]=-\int d^{2}\xi\sqrt{-\gamma}\,(\mu-\frac{J^{2}}{4Ke^{2}})
\label{DS}
\end{equation}
where we have also used (\ref{Jy}). 

Including the usual Maxwell terms, the total action for a
superconducting string is,

\begin{equation}
S=-\frac{1}{4}\int d^{4}x\sqrt{-g}\,F^{\mu\nu}F_{\mu\nu}-
\int d^{2}\xi\,[\sqrt{-\gamma}(\mu-\frac{J^{2}}{4Ke^{2}})+J^{a}A_{a}]
\label{SJ}
\end{equation}

In terms of $y$ the above action is
\cite{witt}\cite{sper}\ldots\cite{linet},

\begin{equation}
S=-\frac{1}{4}\int d^{4}x\sqrt{-g}\,F^{\mu\nu}F_{\mu\nu}-
\int d^{2}\xi\,\{\sqrt{-\gamma}\,[\mu+\frac{1}{2}(\partial y)^{2}]-
\sqrt{2K}e\,\varepsilon^{ab}(\partial_{a}y)A_{b}\}
\label{Sy}
\end{equation}
where $g$ is the determinant of the spacetime metric and we have used
(\ref{Jy}). 

The above action can be recovered equivalently using the fermionic
action \cite{book},

\begin{equation}
\Delta S_{J}=-\int
d^{2}\xi\sqrt{-\gamma}\;(\overline{\Psi}i\!\not\!\!D\Psi)=
-\int d^{2}\xi\sqrt{-\gamma}\,[\frac{1}{2}(\partial y)^{2}-
\sqrt{2K}e\,\varepsilon^{ab}(\partial_{a}y)A_{b}]
\end{equation}
where we have used (\ref{Y}). 

Varying the action with respect to $x^{\mu}$,$A_{\mu}$ and $y$
yields the following equations of motion respectively,

\begin{eqnarray}
 & & [\mu+\frac{1}{2}(\partial y)^{2}]\Box_{2}x^{\mu}-
(\partial^{b}y)(\partial_{b}x^{\mu})\Box_{2}y-
(\partial^{a}y)(\partial^{b}y)(\partial_{a}\partial_{b}x^{\mu})+
J_{\nu}F^{\mu\nu}=0\\
 & & \nonumber\\
 & & \Box_{2}y=\sqrt{2K}e\,\varepsilon^{ab}\partial_{a}A_{b}
\label{boxy}\\
 & & \nonumber\\
 & & \partial_{\mu}F^{\mu\nu}=\Box A^{\nu}=J^{\nu}
\end{eqnarray}
where \mbox{$\Box_{2}\equiv\partial_{a}\partial^{a}$} and we have used
the Coulomb gauge, \mbox{$\partial_{\mu}A^{\mu}=0$} for the Maxwell's
equations. Note also that,
\mbox{$2\varepsilon^{ab}\partial_{a}A_{b}=\varepsilon^{ab}F_{ab}$}. 

Using (\ref{J}) and (\ref{Jy}) the field equation (\ref{boxy}) reduces
to the trivial expression,
\mbox{$\varepsilon^{ab}\partial_{a}\partial_{b}\psi=0$}, which in fact
led to the introduction of $y$ in \cite{witt}. Also, from (\ref{y})
one can easily show that,

\begin{equation}
\Box_{2}\psi=-e\,\partial_{a}A^{a}
\label{boxpsi}
\end{equation}

Using the above and (\ref{J}) one can
obtain the current conservation equation (\ref{Jcons}). 
Employing the Coulomb gauge we obtain,
\mbox{$\partial_{a}\psi=const.$}, i.e. {\em the gradient of the phase
field $\psi$ remains constant 
along the string} \cite{shel}. 

Finally, from (\ref{Jy}) and (\ref{boxy}) we obtain,
\mbox{$\partial_{a}J_{b}=2Ke^{2}F_{ab}$}. Taking the time component of
this, we find,

\begin{equation}
\frac{\partial J}{\partial t}=2Ke^{2}E
\label{JE}
\end{equation}
where \mbox{$E\equiv\varepsilon^{ab}\partial_{a}A_{b}$} is the electric
field, externally applied on the string. 

The above justifies the fact that the strings are considered to be
superconducting. However, as shown in \cite{witt}, (\ref{JE}) breaks
down when the current grows very large. Indeed, there is a maximum
current over which the string loses its
superconducting properties. This occurs when the energy of the current
either permits unwindings of the phase field (bosonic case) or allows
current carriers to escape from the string (fermionic case). In both
cases the maximum current cannot exceed, \mbox{$J_{max}\sim
e\sqrt{\mu}$} \cite{witt}. 

\subsection{Back-reaction and external fields}

The self-inductance of the string may be defined as follows
\cite{hill},

\begin{equation}
L\equiv\frac{1+4Ke^{2}\ln(\Lambda R)}{2Ke^{2}}\simeq 2\ln(\Lambda R)
\label{L}
\end{equation}
where \mbox{$\Lambda^{-1}\sim(\sqrt{\lambda}\eta)^{-1}$} is the string
width and $R$ is a suitable cut-off radius usually taken as the
curvature radius of the string or the inter-string distance of the
string network. In the above \mbox{$\ln(\Lambda R)\simeq 100\gg 1$}. 

Considering self-inductance effects (\ref{JE}) is modified as
\cite{book},

\begin{equation}
\frac{dJ}{dt}=2Ke^{2}(E-L\frac{dJ}{dt})=2K\tilde{e}^{2}E
\end{equation}
where the renormalised charge is,

\begin{equation}
\tilde{e}^{2}\equiv\frac{e^{2}}{1+2Ke^{2}L}\simeq\frac{1}{2KL}
\label{etild}
\end{equation}

Considering back-reaction effects the photon field may be expressed as
\cite{witt}\cite{amst1}, 

\begin{equation}
\hat{A}^{\mu}=
2\int d^{2}\xi\sqrt{-\gamma}\,J^{a}(\xi)\,\partial_{a}x^{\mu}(\xi)\,
\Theta[x^{0}-x^{0}(\xi)]\,\delta([x-x(\xi)]^{2})
\end{equation}
where the step function $\Theta$ signifies the initiation of the
current. 

After some algebra the above reduces to
\cite{witt}\cite{turok}\cite{book}, 

\begin{equation}
\hat{A}^{a}=-2\ln(\Lambda R)J^{a}\simeq -LJ^{a}
\label{AL}
\end{equation}

Using the above, (\ref{Jy}) and (\ref{boxy}) we obtain \cite{book},

\begin{equation}
\Box_{2}\tilde{y}=
\frac{1}{2}\sqrt{2K}\tilde{e}\,\varepsilon_{ab}\hat{F}^{ab}
\label{ytild}
\end{equation}
which is the renormalised version of (\ref{boxy}). In the above
\mbox{$\tilde{y}\equiv(e/\tilde{e})y$} and $\hat{F}^{ab}$ is the
external field strength. The right-hand side of (\ref{ytild}) is the
Lorentz force on the string since 
\mbox{$\varepsilon^{ab}\hat{F}_{ab}=\varepsilon^{0a}[E_{a}+
\varepsilon_{aij}(\partial_{0}x^{i})B^{j}]$}, where $E_{a}$ and $B_{i}$
are the external electric and magnetic fields respectively \cite{book}.

Similarly, for the string current we have
\cite{vil}\ldots\cite{linet}, 

\begin{eqnarray}
J^{\mu} & = &
\int d^{2}\xi\sqrt{-\gamma}\,J^{a}\partial_{a}x^{\mu}(\xi)\,
\delta^{(4)}(x-x(\xi))=\nonumber\\
 & & \nonumber\\
 & = & \int d^{2}\xi\sqrt{-\gamma}\,\varepsilon^{ab}
\partial_{b}y\,\partial_{a}x^{\mu}(\xi)\,
\delta^{(4)}(x-x(\xi))
\end{eqnarray}

From the above, since the quantity $ey$ is unaltered by
renormalisation, we conclude that {\em the string current is
unaffected by charge renormalisation}. Also, (\ref{ytild}) suggests
that, in the absence of external fields
\mbox{$\Box_{2}\tilde{y}=0$} \cite{vil}. Using the renormalised
version of (\ref{Jy}) we can immediately obtain current conservation.

Equivalently, using (\ref{J}) and (\ref{AL}) one can show that,

\begin{equation}
J_{a}=\frac{1}{\tilde{e}L}\partial_{a}\psi
\label{Jpsi}
\end{equation}
which again suggests that the current is conserved since
\mbox{$\partial_{a}\psi=const.$}. 

Finally, we can obtain the renormalised version of the action
by considering the inductance energy of the string \cite{hill},
\mbox{${\cal E}=\frac{1}{2}LJ^{2}$}. Thus, the addition to the
Goto--Nambu action due to the string current is,

\begin{equation}
\Delta S=\frac{1}{2}\int d^{2}\xi\sqrt{-\gamma}LJ^{2}
\end{equation}

Adding the above to (\ref{Sns}) and using also (\ref{etild}) we find,

\begin{equation}
\Delta S=
-\int d^{2}\xi\sqrt{-\gamma}\,(\mu-\frac{J^{2}}{4K\tilde{e}^{2}})
\end{equation}
which is the renormalised analogue of (\ref{DS}).

In a similar way we can regard the total action in (\ref{SJ}) and
(\ref{Sy}) by considering the external fields and the renormalised
values of the physical quantities. From now on, unless stated
otherwise, we will refer to the renormalised values of the latter. For
economy, we will drop the tildes and hats.

\subsection{The string current}

Suppose that the string is lying along the $z$-axis. Then,
\mbox{$J^{\mu}=\delta^{2}(r)J^{\mu}(z,t)$}. With the use of (\ref{Jy})
we find \cite{aryal},

\begin{equation}
J^{\mu}(z,t)=
\sqrt{2K}e\,(\frac{\partial y}{\partial z},0,0,-\frac{\partial
y}{\partial t})
\label{Jzt}
\end{equation}

If \mbox{$\rho_{e}$} is the charge density inside the string then,
\mbox{$\frac{\partial}{\partial t}\rho_{e}\equiv J_{z}$} and 
\mbox{$\frac{\partial}{\partial z}\rho_{e}\equiv J_{0}$}. In view of
(\ref{Jzt}) we can identify,

\begin{equation}
\rho_{e}=-\sqrt{2K}e\,y = \frac{q}{\sqrt{\pi}}\,y
\end{equation}
that is, {\em the field $y$ is a measure of the charge density along
the string} .

Now, the Maxwell's equations give,

\begin{equation}
\frac{1}{r}\frac{\partial}{\partial r}(r\partial_{r}A^{\mu})
=4\pi J^{\mu}\delta^{(2)}(r)\Rightarrow
\partial_{r}A^{\mu}=\frac{2J^{\mu}}{r}
\label{Ar}
\end{equation}

Using the above we can find the electric and magnetic field around the
string,

\begin{eqnarray}
E_{r} & = & \frac{2J^{0}}{r}\nonumber\\
 & & \label{sfields}\\
B_{\theta} & = & \frac{2J^{z}}{r}\nonumber
\end{eqnarray}

The last of the above is the well-known Biot--Savart law for a line
current.

Assuming now that, the charge density along the string is uniformly
distributed, \mbox{$J_{0}=0$} and \mbox{$J^{\mu}=(0,0,0,J)$}. Then,
the electric field vanishes and the only non-zero components of the
electromagnetic field strength are \cite{peter}\cite{linet},

\begin{equation}
F_{Az}=\frac{2J}{r}(\frac{r^{A}}{r})
\label{F}
\end{equation}

\section{The energy--momentum tensor}

\subsection{The string current component}

Using (\ref{Jy}) the action (\ref{DSJ}) may be written as,

\begin{equation}
\Delta S_{J}=\frac{1}{4Ke^{2}}\int d^{2}\xi\sqrt{-\gamma}\,J^{2}
\end{equation}

Varying the above action with respect to the world-sheet metric we
obtain the energy--momentum tensor of the string current 
\cite{turok}\cite{book},

\begin{eqnarray}
\Theta^{ab} & \equiv & \frac{1}{\sqrt{-\gamma}}\frac{\delta\Delta
S_{J}}{\delta\gamma_{ab}}=
\frac{1}{4Ke^{2}}[2J^{a}J^{b}-\gamma^{ab}J^{2}]=\nonumber\\
 & & \nonumber\\
 & = & -\frac{1}{2}(\partial
y)^{2}\gamma_{ab}+(\partial_{a}y)(\partial_{b}y)
\label{Theta}
\end{eqnarray}

It can be easily checked that the above tensor is traceless. Using
(\ref{ytild}) we find,

\begin{equation}
\partial_{b}\Theta^{b}_{a}=\partial_{a}y\Box_{2}y=
\frac{1}{2}J^{\mu}F_{\mu\nu}\partial_{a}x^{\nu}
\end{equation}

In the absence of external fields the above becomes, 
\mbox{$\partial_{b}\Theta^{b}_{a}=0$}. In view of (\ref{Theta}) this
gives,

\begin{eqnarray}
{\bf J}\cdot{\bf E}=0 & \mbox{and}  & {\bf J}\times{\bf B}=0
\end{eqnarray}
which are obviously satisfied by the self-fields (\ref{sfields}) of the
string. 

\subsection{With no electromagnetic back-reaction}

From (\ref{Sy}) the energy--momentum tensor of the string is given by
\cite{book}, 

\begin{eqnarray}
T^{ab} & = & \frac{1}{\sqrt{-\gamma}}
\frac{\delta
S}{\delta\gamma^{ab}}=-\gamma^{ab}[\mu+\frac{1}{2}(\partial y)^{2}]+
\partial_{a}y\partial_{b}y\Rightarrow\nonumber\\
 & & \label{T1a}\\
T^{\mu\nu} & = & -(\mu\gamma^{ab}-\Theta^{ab})\,
\partial_{a}x^{\mu}\partial_{b}x^{\nu}\nonumber
\end{eqnarray}

Using (\ref{Theta}) with \mbox{$J^{\mu}\equiv(0,0,0,J)$} it is easy to
see that \cite{linet},

\begin{equation}
\Theta_{0}^{0}=-\Theta_{z}^{z}=-\frac{J^{2}}{4Ke^{2}}
\end{equation}

Thus, we can write (\ref{T1a}) as \cite{turok},

\begin{equation}
T^{\mu}_{\nu}=-\delta^{(2)}(r)\,
\mbox{diag}(\mu+\frac{J^{2}}{4Ke^{2}},0,0,\mu-\frac{J^{2}}{4Ke^{2}})
\label{T1}
\end{equation}

Note that the above is a core solution which does not include
electromagnetic back-reaction effects. 

\subsection{With electromagnetic back-reaction}

The electromagnetic energy--momentum tensor is,

\begin{equation}
T^{\mu\nu}_{em}=F^{\mu}_{\lambda}F^{\nu\lambda}-
\frac{1}{4}g^{\mu\nu}F_{\lambda\rho}F^{\lambda\rho}
\label{Tem}
\end{equation}
where $F^{\mu\nu}$ refers to the self-fields of the string.

Using (\ref{F}) one easily finds,

\begin{eqnarray}
T^{00}_{em} & = & T^{zz}_{em}=\frac{2J^{2}}{r^{2}}\nonumber\\
 & & \label{T2}\\
T^{AB}_{em} & = & \frac{2J^{2}}{r^{2}}\,
(\frac{2r^{A}r^{B}}{r^{2}}-\delta^{AB})\nonumber
\end{eqnarray}

With a suitable rotation of the $r^{A}$ coordinates (\ref{T2}) 
can be written as \cite{HK}\cite{babul}, 

\begin{equation}
[T_{em}(r)]^{\mu}_{\nu}=\frac{2J^{2}}{r^{2}}\,\mbox{diag}(-1,1,-1,1)
\end{equation}

The above tensor is traceless but not well-defined at
\mbox{$r\rightarrow 0$}. In order to overcome this problem we employ a
method introduced by Linet \cite{linet} and expand the components of
$T^{\mu\nu}_{em}$ as distributions around the string core 
(see also \cite{peter}). We use the fact that
\mbox{$2/r^{2}=\Delta_{2}[\ln(r/r_{0})]^{2}$} 
where \mbox{$\Delta_{2}\equiv
(1/r)\partial_{r}(r\partial_{r})$} is the 2-dimensional Laplacian and
$r_{0}$ is the radius of the string core.\footnote{The core radius is
defined as the distance where the string gauge fields assume their
long-range logarithmic behaviour whereas the rest of the vortex fields
become negligible. The above results are also in agreement with
\cite{HK} if one takes into account that
\mbox{$A(r_{0})=-LJ=-J/2Ke^{2}$}.} The energy--momentum tensor
(\ref{T2}) is, then, written as, 

\begin{eqnarray}
T^{00}_{em} & = & T^{zz}_{em}=J^{2}\Delta_{2}
[\ln(r/r_{0})]^{2}\nonumber\\
 & & \label{Tln}\\
T^{AB}_{em} & = & -2J^{2}\partial_{A}\partial_{B}[\ln(r/r_{0})]
\nonumber
\end{eqnarray}

However, since 
\mbox{$\Delta_{2}\ln(r/r_{0})=\frac{1}{2}\delta^{(2)}(r)$}, one finds,
\mbox{$\partial_{A}T^{AB}_{em}=-J^{2}\partial^{B}\delta^{(2)}(r)$} and
 $T_{em}^{\mu\nu}$ is not conserved. In order to have conservation we
need to add the term \mbox{$J^{2}\delta^{AB}\delta^{(2)}(r)$} on
$T^{AB}_{em}$. The addition of this term suggests that the core
solution (\ref{T1}) for the string energy--momentum tensor 
becomes \cite{MP}\cite{linet},

\begin{equation}
T^{\mu}_{\nu}=-\delta^{(2)}(r)\,
\mbox{diag}(\mu+\frac{J^{2}}{4Ke^{2}},-\frac{J^{2}}{2},
-\frac{J^{2}}{2},\mu-\frac{J^{2}}{4Ke^{2}})
\label{T}
\end{equation}

The above is a core solution that takes into account electromagnetic
back-reaction. As can be seen in (\ref{T}), the string current
increases the linear mass density $\mu$ of the string by a factor
$\frac{1}{2}LJ^{2}$ which corresponds to the self-inductance energy
density of the current. Also, it generates pressure due to the
inertia of the charge carriers \cite{babul}, both towards the
perpendicular direction ($T^{1}_{1}$ and $T^{2}_{2}$ terms) and along
the string. The latter also reduces the total string tension $T^{z}_{z}$.

In fact, (\ref{T}) suggests that, for high enough currents, the string
tension may be diminished to zero. In that case the strings would not
have a driving force to untangle. Also, string loops would not
collapse, but remain stable and form the so called ``springs''
\cite{turok}\cite{hill}\cite{spring}. However, it has been argued
\cite{shel} that the back-reaction of the current would shrink the
$\sigma$-condensate in the string core (in the bosonic case) thus
reducing the pressure from the charge carriers that counteracts the
string tension. In overall, the string tension may not decrease as
rapidly due to the current as implied by (\ref{T}). Indeed, numerical
simulations \cite{peter2} have shown that the string tension is little
affected by changes of the string current. It would be more accurate,
then, to express the core solution as,

\begin{equation}
T^{\mu}_{\nu}=-\delta^{(2)}(r)\,\mbox{diag}(U,W,W,T)
\label{TU}
\end{equation}
where \mbox{$W=-J^{2}/2$} \cite{peter}. 
Then the overall solution for the string energy--momentum tensor may
be written as,

\begin{eqnarray}
T^{00}_{em} & = & U\delta^{(2)}(r)+J^{2}\Delta_{2}
[\ln(r/r_{0})]^{2}\nonumber\\
 & & \nonumber\\
T^{zz}_{em} & = & -T\delta^{(2)}(r)+J^{2}\Delta_{2}
[\ln(r/r_{0})]^{2}\label{Ts}\\
 & & \nonumber\\
T^{AB}_{em} & = & J^{2}\delta^{AB}\delta^{(2)}(r)
-2J^{2}\partial_{A}\partial_{B}[\ln(r/r_{0})]
\nonumber
\end{eqnarray}

Thus, the following should be regarded as extreme estimates,

\begin{eqnarray}
U & \simeq & \mu+\frac{J^{2}}{4Ke^{2}}\label{U}\\
 & &\nonumber\\
T & \simeq & \mu-\frac{J^{2}}{4Ke^{2}}\label{TT}
\end{eqnarray}

\section{The string spacetime}

\subsection{The string metric}

Using (\ref{Ts}) one can solve the Einstein equations,

\begin{equation}
R^{\mu\nu}-\frac{1}{2}g^{\mu\nu}R^{\lambda}_{\lambda}=8\pi GT^{\mu\nu}
\end{equation}
where \mbox{$G=m_{P}^{-2}$} is Newton's gravitational constant.

In first order in $G$ the metric is found to be \cite{HK}\cite{peter},

\begin{eqnarray}
ds^{2} & = &
\{1+4G[J^{2}+(U-T)]\ln(r/r_{0})+4GJ^{2}[\ln(r/r_{0})]^{2}\}
(-dt^{2}+dr^{2})\nonumber\\
 & & \nonumber\\
 & + & \{1-8G(U+\frac{J^{2}}{2})-
4G[J^{2}-(U-T)]\ln(r/r_{0})+4GJ^{2}[\ln(r/r_{0})]^{2}\}
r^{2}d\theta^{2}\nonumber\\
 & & \nonumber\\
 & + &
\{1+4G[J^{2}-(U-T)]\ln(r/r_{0})-4GJ^{2}[\ln(r/r_{0})]^{2}\}dz^{2}
\label{ds}
\end{eqnarray}
where \mbox{$J=2W$}.

The metric of the spacetime perpendicular to the string may be found
by setting, \mbox{$dz=0$}. Then the above gives,

\begin{equation}
ds^{2}_{\perp}=(1-h_{00})[-dt^{2}+dr^{2}+(1-b)r^{2}d\theta^{2}]
\label{dsT}
\end{equation}

where,

\begin{equation}
h_{00}=-4G[J^{2}+(U-T)]\ln(r/r_{0})-4GJ^{2}[\ln(r/r_{0})]^{2}
\label{h00}
\end{equation}

and to first order in $G$,

\begin{eqnarray}
1-b & = & \frac{1-8G(U+J^{2}/2)-
4G[J^{2}-(U-T)]\ln(r/r_{0})+4GJ^{2}[\ln(r/r_{0})]^{2}}{
1+4G[J^{2}+(U-T)]\ln(r/r_{0})+4GJ^{2}[\ln(r/r_{0})]^{2}}\nonumber\\
 & & \\
 & \simeq & 1-8G(U+\frac{J^{2}}{2})-8GJ^{2}\ln(r/r_{0})\nonumber
\end{eqnarray}

From (\ref{dsT}) we see that the spacetime around a superconducting
string resembles the conical spacetime of a non-superconducting
string. Indeed, if we set \mbox{$J=0$} and \mbox{$U=T=\mu$}, then
\mbox{$h_{00}=0$} and \mbox{$(1-b)=(1-8G\mu)$}, and (\ref{dsT})
reduces to the non-superconducting string, conical--spacetime solution
with deficit angle \mbox{$\delta=8\pi G\mu$}. In the case of
non-vanishing current, though, the deficit angle is,

\begin{equation}
\delta=b\pi=8\pi G\{U+J^{2}[\frac{1}{2}+\ln(r/r_{0})]\}
\label{delr}
\end{equation}

Thus, the spacetime is not exactly conical since $\delta$ is dependent
on $r$. However, this logarithmic dependence is very weak and, if one
is interested in astrophysical effects (such as primordial magnetic
field generation) then the logarithmic dependence may well be
approximated as,

\begin{equation}
\ln(r/r_{0})\simeq\ln(\Lambda R)
\label{app}
\end{equation}

Under this approximation (\ref{delr}) is in good agreement with the
more rigourous calculations of \cite{HK}\cite{peter}. Now, using
(\ref{U}) and (\ref{etild}) we find,

\begin{equation}
\delta\simeq 8\pi G[\mu+\frac{1}{2}(QJ)^{2}]
\label{d}
\end{equation}

where,

\begin{equation}
Q\equiv\sqrt{1+4\ln(\Lambda R)}\simeq 20
\label{Q}
\end{equation}

\subsection{The gravitational field}

From (\ref{ds}) it can be seen that the string metric deviations from
Minkowski spacetime are of the order of \mbox{$G\mu\leq
10^{-6}$}. Thus, we can use linear theory to approximate the
gravitational field. The metric, then, can be written as,

\begin{equation}
g_{\mu\nu}=\eta_{\mu\nu}+h_{\mu\nu}
\end{equation}
where \mbox{$\eta_{\mu\nu}=(-1,1,1,1)$} is the Minkowski metric. 

The geodesic equation is, \mbox{$\frac{du^{\lambda}}{d\tau^{2}}+
\Gamma^{\lambda}_{\mu\nu}u^{\mu}u^{\nu}=0$} where, 
\mbox{$u^{\mu}=\frac{dx^{\mu}}{d\tau}\simeq (1,{\bf v})$} 
is the 4-velocity and \mbox{$\Gamma^{\lambda}_{\mu\nu}
\simeq\frac{1}{2}\eta^{\lambda\rho}(\partial_{\nu}h_{\mu\rho}+
\partial_{\mu}h_{\nu\rho}-\partial_{\rho}h_{\mu\nu})$} are the
Christoffel symbols in the linear approximation. Since, \mbox{$|{\bf
v}|\ll 1$} the geodesic equation becomes,

\begin{equation}
\frac{d^{2}x^{i}}{d\tau^{2}}+\Gamma^{i}_{00}=0
\label{geo}
\end{equation}
where $i$ denotes the spacial coordinates and \mbox{$\tau\simeq t$} is
the proper time. Since, 
\mbox{$\Gamma^{i}_{00}=-\frac{1}{2}\partial_{i}h_{00}$} we find,

\begin{equation}
{\bf f}=\frac{1}{2}\nabla h_{00}
\end{equation}

Inserting (\ref{h00}) into the above we obtain, 

\begin{equation}
f(r)=-\frac{2GJ^{2}}{r}\,[1+\frac{U-T}{J^{2}}+2\ln(r/r_{0})]
\label{fF}
\end{equation}

Using (\ref{etild}), (\ref{U}), (\ref{TT}) and the approximation
(\ref{app}) we find, 

\begin{equation}
f(r)\simeq -\frac{2G(QJ)^{2}}{r}
\label{f}
\end{equation}

The above is an attractive gravitational force similar to the one of a
massive rod of linear mass density \mbox{$\sim (QJ)^{2}$}.

In (\ref{f}) we have used the extreme values (\ref{U}) and (\ref{TT})
of the string linear mass density and tension. As mentioned, though,
the string tension may be larger than suggested by (\ref{TT}). In that
case $Q$  in the expression (\ref{f}) would decrease 
and the gravitational pull of the string will be weakened since it
depends on the difference \mbox{$(U-T)$} as shown in (\ref{fF}). 
However,  (\ref{ff}) suggests that, even in the extreme case, when
\mbox{$U=T$}, the attractive force is decreased only by a factor of 2. 

\subsection{Particle deflection}

Writing the metric (\ref{dsT}) in cartesian coordinates we have,

\begin{equation}
ds^{2}_{\perp}=(1-h_{00})(-dt^{2}+dx_{k}dx^{k})
\end{equation}
where the \mbox{$k=1,2$} and
we have assumed that the string lies on the $z$-axis. Note that we
need to extract from the above a wedge of angle width $\delta$
\cite{tvach}. Then,

\begin{equation}
d\tau^{2}=-d^{2}s_{\perp}=(1-h_{00})dt^{2}(1-\dot{x}_{k}\dot{x}^{k})
\end{equation}
where the dots signify derivation with respect to time.

Using \mbox{$h_{00}\ll 1$} and 
\mbox{$\Gamma_{00}^{i}=-\frac{1}{2}\partial_{i}h_{00}$}, we insert the
above to the geodesic equation (\ref{geo}) and obtain \cite{tvach},

\begin{equation}
2\ddot{x}^{i}=(1-\dot{x}_{k}\dot{x}^{k})\partial^{i}h_{00}
\label{acl}
\end{equation}
where \mbox{$i=1,2$}.

The above gives the acceleration felt by the particles due to the
gravitational pull of the string field, in the frame of the string. 

Suppose that the string moves in the $x$-direction with constant
velocity \mbox{$-v=-\sqrt{\dot{x}_{k}\dot{x}^{k}}$}. Then for a
particle in the position \mbox{$(x,y)$} we have initially,
\mbox{$x=vt$} and \mbox{$y=const.$}. The velocity boost felt by the
particle towards the $y$-direction after its encounter with the string
is \cite{book},

\begin{equation}
v_{y}=\int_{-\infty}^{\infty}\ddot{y}dt=
\frac{1}{2v\gamma^{2}}\int_{-\infty}^{\infty}\partial_{y}h_{00}dx
\end{equation}
where \mbox{$\gamma^{-1}\equiv\sqrt{1-v^{2}}$} is the Lorentz factor
and we have also used (\ref{acl}).

Using (\ref{h00}) and (\ref{app}) the above gives,

\begin{equation}
v_{y}=-\frac{2\pi G(QJ)^{2}}{v\gamma^{2}}
\end{equation}

Taking into account the deficit angle, the relative boost between two
particles on the opposite sides of the string is, 
\mbox{$\Delta v_{y}=v\delta+2|v_{y}|$}. Switching to the particle frame
gives,

\begin{equation}
u=\gamma\Delta v_{y}=8\pi G\mu v\gamma+4\pi
G(QJ)^{2}(v\gamma+\frac{1}{v\gamma})
\label{boost}
\end{equation}
where we have also used (\ref{d}). 

\subsection{Deceleration of the string}

Avelino  and Shellard \cite{paper} have realised that the
deflection of particles in the string spacetime results into a net
drag of the plasma behind the string. This is due to the fact that the
magnitude of the particle velocity is unaltered after its interaction
with the string. The back-reaction of this effect is a decelerating
force on the string. The velocity of plasma dragging may be estimated
by Taylor expansion of the particle velocity. In the lowest order in
$G$ we find,

\begin{equation}
\delta v_{x}\simeq-\frac{1}{2}\frac{u^{2}}{v}=
-32\pi^{2}G^{2}\mu v\gamma^{2}[\mu+J^{2}(1+\frac{1}{v^{2}\gamma^{2}})]-
8\pi^{2}G^{2}(QJ)^{4}v\gamma^{2}(1+\frac{1}{v^{2}\gamma^{2}})
\label{vx}
\end{equation}

The momentum change of the string is, 
\mbox{$dp=(\rho\delta v_{x})dx\,dy\,dz$}, where \mbox{$\rho=3/32\pi
Gt^{2}$} is the energy density of the universe with
\mbox{$\Omega=1$}. Thus, the drag force per unit length on the string
is, 

\begin{equation}
f_{x}=\int\frac{d^{2}p}{dtdz}\,dxdy=2R\rho v\,\delta v_{x}
\end{equation}
where \mbox{$R\sim vt$} is the inter-string distance of the network
\cite{mine}. 

Inserting (\ref{vx}) into the above we find,

\begin{equation}
f_{x}=-6\pi GHv\{2\mu^{2}v^{2}\gamma^{2}+
\frac{1}{2}(QJ)^{4}(v\gamma+\frac{1}{v\gamma})^{2}+
2\mu J^{2}(v^{2}\gamma^{2}+1)\}
\label{fx}
\end{equation}
where \mbox{$H^{-1}=2t$} is the Hubble radius.

The above force should be compared with the plasma friction force on
the strings, \mbox{$f_{s}=\max\{\rho v/m,Jv\sqrt{\rho}\}$},
\cite{mine} where $m$ is the particle mass. It is easy to show that
throughout the range of $J$ the above drag force is always subdominant
and, thus, it does not affect the network evolution.\footnote{For
example, in the case of non-superconducting strings (\mbox{$J=0$}) the
drag force dominates string friction at the temperature,
\mbox{$T_{f}\sim (G\mu)^{2}m_{P}$}. Comparing this with the
temperature \mbox{$T_{*}\sim(G\mu)m_{P}$} when the network reaches the
horizon scaling solution \cite{mine}, we see that $f_{x}$ has no
effect during the friction era of the network. During horizon scaling
we can compare $f_{x}$ with the string tension $\sim(\mu/t)$ and
verify that the latter dominates at all times.}

\section{Primordial magnetic field generation}

Although the motion of the strings is not significantly affected by
the dragging of the plasma, the latter may gain substantial momentum
during this process. Such momentum may introduce turbulence which
could generate a primordial magnetic field.

\subsection{Momentum transfer}

In the one scale model the scale of the string network inter-string
distance is comparable to the string curvature radius, \mbox{$R\sim
vt$}. A string segment of length $\sim R$ may transfer momentum to the
plasma contained in an inter-string volume $\sim R^{3}$. In that way
the network may introduce vortical motions to the plasma on
inter-string scales. An estimate of the plasma
rotational velocity $v_{rot}$ may be obtained as follows,

\begin{equation}
FR \simeq \frac{1}{2}Mv_{rot}^{2}
\end{equation}
where \mbox{$M\sim\rho R^{3}$} is the mass of an inter-string volume
and $F$ is the total force on the plasma by a string segment of length
$R$. From the above the rotational velocity is estimated as,

\begin{equation}
v_{rot}\sim\frac{\sqrt{GF}}{v}
\label{vrot}
\end{equation}

The total force may be obtained from (\ref{fx}) as follows,

\begin{eqnarray}
F & = & \int_{0}^{R}f_{x}dz=vtf_{x}\Rightarrow\nonumber\\
 & & \label{ff}\\
F & = & 3\pi Gv^{2}[2\mu^{2}v^{2}\gamma^{2}+
\frac{1}{2}(QJ)^{4}(v\gamma+\frac{1}{v\gamma})^{2}+
2\mu J^{2}(v^{2}\gamma^{2}+1)]\nonumber
\end{eqnarray}

As shown in \cite{mine}, for currents greater than a critical value
$J_{c}$ the network can never exit the friction
domination era of its evolution. In \cite{mine} $J_{c}$ is estimated
as, 

\begin{equation}
J_{c}\sim\sqrt{G}\,\mu 
\label{Jc}
\end{equation}

However, even for such strong currents the network does reach a
scaling solution during which the strings move with a terminal
velocity given by \cite{mine},

\begin{equation}
v_{T}^{2}\sim\frac{G\mu}{\sqrt{GJ^{2}}}
\label{vT}
\end{equation}

As implied in \cite{mine} the string current is influential on the
behaviour of the network only  if \mbox{$J\geq J_{c}$}. Thus, we will
consider this case only. Using (\ref{vT}) the total force (\ref{ff})
becomes,

\begin{equation}
F=3\pi[2(G\mu)^{2}(\frac{\mu}{J^{2}})\mu+
\frac{1}{2}Q^{4}(G\mu)^{2}J^{2}+
\frac{1}{2}Q^{4}(GJ^{2})J^{2}+
Q^{4}(GJ^{2})\sqrt{G\mu}\sqrt{\frac{\mu}{J^{2}}}J^{2}+
2(G\mu)\mu^{2}+2\sqrt{GJ^{2}}\mu^{2}]
\end{equation}
In the above the 2nd, the 4th and the 5th terms of the right hand side
may be disregarded by means of the condition
\mbox{$J>J_{c}$}. By comparing the remaining terms with each other it
can be shown that the last term remains also always
subdominant. Therefore, the above can be written as,

\begin{equation}
F=6\pi J^{2}[(G\mu)^{2}(\frac{\mu}{J^{2}})^{2}+
\frac{1}{2}Q^{4}(GJ^{2})]
\label{FF}
\end{equation}

Considering the overall range of $J$ we can rewrite the above
as,\footnote{Note that \mbox{$J<J_{c}\Rightarrow v=1$}.}  

\begin{equation}
F\simeq\left\{
\begin{array}{lr}
3\pi G(QJ)^{4} & \;\;\;\;\;J_{c1}<J\leq J_{max}\\
 & \\
6\pi(G\mu)^{2}(\mu/J^{2})\mu & \;\;\;\;\;J_{c}<J\leq J_{c1}\\
 & \\
6\pi G\mu^{2} & \;\;\;\;\;J\leq J_{c} 
\end{array}
\right.
\label{F3}
\end{equation}
where the critical current $J_{c1}$ is,

\begin{equation}
J_{c1}\equiv Q^{-2/3}(G\mu)^{1/6}\sqrt{\mu}
\label{Jc1}
\end{equation}

For currents stronger that the above value the current dependent term
in (\ref{boost}) becomes dominant, i.e. the gravitational pull of the
string is actually felt by the particles. 

Inserting (\ref{F3}) into (\ref{vrot})  and using also (\ref{vT}) we
find,

\begin{equation}
v_{rot}\simeq\left\{
\begin{array}{lr}
\sqrt{3\pi}Q^{2}(G\mu)^{-1/4}(GJ^{2})
(\mu/J^{2})^{-1/4} & \;\;\;\;\;J_{c1}<J\leq J_{max}\\
 & \\
\sqrt{6\pi}(G\mu)^{5/4}(\mu/J^{2})^{1/4} & \;\;\;\;\;J_{c}<J\leq J_{c1}\\
 & \\
\sqrt{6\pi}(G\mu) & \;\;\;\;\;J\leq J_{c} 
\end{array}
\right.
\label{vrot1}
\end{equation}

\subsection{The Harrison--Rees magnetic field}

Harrison \cite{hary} suggested first that turbulence in an expanding
universe may generate a magnetic field since the turbulent velocity
of the plasma would be different for the ions and the
electrons. 

Consider a rotating volume $V$ of plasma. Suppose that the angular
velocities $\omega_{m}$ and $\omega_{r}$ of the ion and the electron
fluid respectively are uniform inside $V$. Then, since
\mbox{$V\propto a^{-3}$}, we find that,

\begin{eqnarray}
\rho_{m}V=const.\;\;\;\;\; & \mbox{and} & 
\;\;\;\;\;\rho_{r}V^{4/3}=const.
\end{eqnarray}
where \mbox{$\rho_{m}\propto a^{-3}$} is the ion density, which scales
as pressureless matter, and \mbox{$\rho_{r}\propto a^{-4}$} is the
electron density, which scales as radiation due to the strong coupling
between the electrons and the photons through Thompson scattering. The
angular momentum \mbox{$I=\rho\omega V^{5/3}$} of each plasma
component has to be conserved. This suggests that,

\begin{eqnarray}
\omega_{m}\propto V^{-2/3}\propto a^{-2}\;\;\;\;\; & 
\mbox{and} & \;\;\;\;\;\omega_{r}\propto V^{-1/3}\propto a^{-1}
\end{eqnarray}

Thus, the ion fluid spins down faster than the electron-photon gas. 
Consequently, a current is generated which creates a magnetic field in
the volume $V$. 

Rees, however, has shown that expanding volumes of spinning plasma are
unstable and decay with cosmic expansion compared to irrotational
density perturbations \cite{rees}. He suggested instead a different
version of vortical magnetic field generation involving the scattering
of the electrons on the microwave background radiation. This would
tend to damp the vortical motions of the electrons in contrast to the
ions which would stay unaffected. The result is again the generation of
currents but, this time, it is the electron fluid that slows down. The
mechanism applies to gravitationally bound spinning bodies such as
those formed when structure formation begins. 

In both cases, the Maxwell's equations suggest \cite{hary}, 

\begin{equation}
{\bf B}\simeq-\frac{m}{e}\,{\bf w}
\end{equation}

where \mbox{$m\sim 1\,GeV$} is the nucleon mass and {\bf w} is the
vorticity given by, 

\begin{equation}
{\bf w}=\nabla\times{\bf v}_{rot}
\end{equation}

From the above the magnetic field generated over inter-string distances
is, 

\begin{equation}
B\simeq\frac{m}{e}(\frac{v_{rot}}{R})
\label{Bv}
\end{equation}

The turbulence inside the wake of cosmic strings is expected to ionise
the plasma even after decoupling \cite{book}. Also for superconducting
cosmic strings the existence of a magnetocylinder around the core
\cite{mine} is expected to induce further charge separation due to the
difference of the inertia of the scattered particles. 
Since \mbox{$R\propto t$}, from (\ref{Bv}) it is evident that the
stronger field will be generated at early times. Thus, we will
calculate the magnetic field generated at 
\mbox{$t_{eq}\sim 10^{11}sec$}, the time of equal
matter and radiation densities, since this is the earliest that large
scale streaming of the plasma is possible. From (\ref{vrot1}) and
(\ref{B}) we obtain,

\begin{equation}
B_{eq}\simeq\frac{\sqrt{6\pi}\,m}{et_{eq}}\times\left\{
\begin{array}{lr}
\frac{1}{\sqrt{2}}Q^{2}(G\mu)^{-1/2}(GJ^{2})
(\mu/J^{2})^{-1/2} & \;\;\;\;\;J_{c1}<J\leq J_{max}\\
 & \\
G\mu & \;\;\;\;\;J\leq J_{c1} 
\end{array}
\right.
\label{Beq}
\end{equation}

For \mbox{$J=J_{max}\simeq\sqrt{\mu}$} the above gives,

\begin{equation}
B^{max}_{eq}\sim 10^{-13}\sqrt{G\mu}\,Gauss
\end{equation}

which is coherent over comoving scales,

\begin{equation}
l\sim(v_{T}t_{eq})(\frac{t_{pr}}{t_{eq}})^{2/3}\sim
10^{2}(G\mu)^{1/4}M\!pc
\end{equation}
where \mbox{$t_{pr}\sim 10^{18}sec$} is the present time.

For energy scales corresponding to a grand unified theory (GUT) phase
transition, \mbox{$G\mu\sim 10^{-6}$} and the above suggest that
\mbox{$B^{max}_{eq}\sim 10^{-16}Gauss$} with a coherency of the order
of \mbox{$l\sim 1\,M\!pc$} which is quite sufficient to seed the
galactic magnetic fields. However, as we discuss below, this estimate
may be over-optimistic.

\subsection{Temperature anisotropy constraint}

The deficit angle of the string spacetime apart from deflecting the
trajectories of particles affects light propagation as well
\cite{book}. From (\ref{dsT}) by setting \mbox{$ds=0$} we see that
photons are boosted by the deficit angle even though the prefactor
\mbox{$(1-h_{00})$} is irrelevant. As a result a string moving in
front of radiation will blueshift light due to the Doppler
effect. Thus, a string network is expected to generate temperature
anisotropies on the microwave background radiation. These cannot
exceed the observed values, 

\begin{equation}
(\frac{\Delta T}{T})_{_{rms}}\leq 10^{-6}
\end{equation}

The anisotropy generated by a single string is \cite{book},

\begin{equation}
(\frac{\Delta T}{T})_{_{S}}\sim\delta v\gamma
\end{equation}

The rms anisotropy due to a network of cosmic strings is estimated as
\cite{mine}\cite{leandros},

\begin{equation}
(\frac{\Delta T}{T})_{_{rms}}\sim\frac{H^{-1}}{R}(\frac{\Delta T}{T})_{S}
\end{equation}

From the above and (\ref{d}) we find that for superconducting cosmic
strings,

\begin{equation}
(\frac{\Delta T}{T})_{_{rms}}\sim\delta\sim G[\mu+\frac{1}{2}(QJ)^{2}]
\label{Trms}
\end{equation}

Thus, for \mbox{$G\mu\leq 10^{-8}$} even the
maximum current could not challenge the observations. However, for GUT
strings one cannot allow the current to reach its maximum value. 
Indeed, there is some doubt whether
superconducting strings may attain the maximum
current \cite{babul} since the energy density of the
current \mbox{$\frac{1}{2}LJ^{2}\simeq\ln(\Lambda R)J^{2}$} should
not exceed the string linear mass density $\mu$ \cite{aryal}. This
implies that the current term in (\ref{Trms}) should always remain
subdominant. Therefore,

\begin{equation}
J\leq J_{c2}\equiv\frac{\sqrt{\mu}}{Q}
\label{Jc2}
\end{equation}

By evaluating the (\ref{Beq}) with \mbox{$J=J_{c2}$} we obtain the
maximum permissible magnetic field for GUT energy scales, 

\begin{equation}
B^{max}_{eq}\sim 10^{-15}Q^{-1}\sqrt{G\mu}\,Gauss\sim 10^{-19}Gauss
\end{equation}
with coherence length,

\begin{equation}
l\sim 10^{2}\sqrt{Q}(G\mu)^{1/4}M\!pc\sim 1\,M\!pc
\end{equation}

The above field is of sufficient strength and coherency to seed the
galactic dynamo process and generate the observed galactic magnetic
fields. 

\section{Discussion and conclusions}

We have showen that GUT superconducting cosmic strings are able to
generate turbulence on inter-string scales, which gives rise to a
primordial magnetic field of enough strength to seed the dynamo
process in galaxies and account for the observed galactic magnetic
fields. Moreover, the generated field is coherent over very large
scales, comparable with the protogalactic ones before gravitational
collapse commenced. Turbulence and coherent rotation on these scales
may also be related to the fragmentation process
of galaxy formation. Furthermore, the existence of a primordial field,
coherent over an entire protogalaxy, may have played a crucial role in
removing angular momentum in a similar way as in the case of the
collapse of protostellar clouds \cite{zeld}\cite{mestel}. 

Due to excessive microwave temperature anisotropies, a maximum string
current may not be acceptable. However, even under this constraint, 
the magnetic field generated is still adequate to seed the galactic
dynamo. A possibly
stronger constraint may be implemented by considering the density
inhomogeneities due to the string wakes, since their growth is
enhanced not only due to the deficit angle but also, due to the
gravitational pull of the strings \cite{book}. However, the magnitude
of the overdensities is strongly dependent on the dark matter model
used. Indeed, string wake overdensities may be substantially
suppressed in the case of hot dark matter, due to free streaming
\cite{book}. 

In our estimates of the generated magnetic field we have used the
extreme values (\ref{U}) and (\ref{TT}) for the string linear mass
density and tension. As we mentioned though, the string tension may
be larger than suggested by (\ref{TT}). This
however, would relax the temperature anisotropy constraint and the
density inhomogeneity constraint even more. 

We have implicitly assumed that the string current remains constant
during the network evolution. Indeed, the current is dynamically
conserved as shown by (\ref{Jcons}). Also, in the bosonic case
(\ref{Jpsi}) suggests that current conservation is ensured on
topological grounds. However, most of the field theory of Section 2,
concerns a straight and infinite superconducting string. A realistic
string may be much more complicated. Still, it can be shown
\cite{mine} that even in this case the string current remains constant
when the network reaches a scaling solution.

Our treatment is similar to this of Avelino and Shellard \cite{paper}.
In their work, however, they consider the case of wiggly strings,
which also have an attractive gravitational field due to the
difference between their renormalised linear mass density and
tension. The latter is a result of the tangled shape of wiggly strings,
which, in effect, increases the string length (and thus, the linear
mass density) between two fixed points on the string while also
decreasing the tension due to the random orientation of the string
microstructure \cite{wiggly}. For superconducting strings,
microstructure is suppressed by electromagnetic radiation emission
\cite{grays} and the difference between the linear mass density and
tension is due to the existence of a current, which increases the
energy content of the string while generating pressure 
and reducing, thus, the string tension \cite{babul}. Therefore, the
gravitational field of superconducting strings arises in a
qualitatively different way that the one of wiggly strings. Moreover,
there are also quantative differences between the two cases, since,
for strong enough currents, a superconducting string network reaches a
different scaling solution, due to excessive plasma friction
\cite{mine}. As a result, the inter-string distance is much smaller
than the horizon and the strings are slow moving. The above enable
superconducting strings to generate much stronger magnetic fields than
wiggly strings. In fact, since the string microstructure is suppressed
by the existence of a current, wiggly strings may be regarded as the
limit of superconducting strings when the current is very small.

The superconducting string model provides a realistic mechanism for
primordial magnetic field generation. The field generated by
natural (GUT-scale) values of the parameters is coherent over
protogalactic scale and strong enough to seed the dynamo process in
galaxies. Thus, our mechanism may be considered as a prime candidate
to explain the observed galactic magnetic fields. 

\bigskip

\bigskip

{\noindent{\Large{\bf Acknowledgements}}}

\nopagebreak[4]

\bigskip

\nopagebreak[4]

I would like to thank A.C. Davis for discussions and her support
during the first stages of this work and M. Papastathi for her
edditing help. This work was supported by the Isaac Newton Fund
(Trinity College, Cambridge), and the Greek State Scholarships
Foundation (I.K.Y.).

\end{document}